\documentclass{ifacconf}

\usepackage{graphicx}      
\usepackage{natbib}        
\usepackage{amssymb}

\usepackage{multirow} 
\usepackage{natbib}
\usepackage{algorithm}
\usepackage[noend]{algpseudocode}

\usepackage[flushleft]{threeparttable}

\algrenewcommand\algorithmicrequire{\textbf{Input:}}
\algrenewcommand\algorithmicensure{\textbf{Output:}}

\usepackage[acronym, toc]{glossaries}

\begin{document}
\begin{frontmatter}
\newcommand{\pnha}[1]{{\color{red}{NH: #1}}}
\newcommand{\dimpa}[1]{{\color{blue}{DP: #1}}}
\newcommand{\roga}[1]{{\color{green}{RG: #1}}}
\newcommand{\yaqpr}[1]{{\color{orange}{YP: #1}}}

\newcommand{\rebuttal}[1]{{\color{red}{#1}}}
\newcommand{\DTU}[1]{#1$^1$}
\newcommand{\SIMAC}[1]{#1$^2$}

\newcommand{\figref}[1]{Fig.~\ref{#1}}

\providecommand{\R}{\ensuremath \mathbb{R}}
\providecommand{\N}{\ensuremath \mathbb{N}}
\providecommand{\Z}{\ensuremath \mathbb{Z}}
\newcommand{\SE}{\regtext{SE}}
\newcommand{\SO}{\regtext{SO}}

\newcommand{\regtext}[1]{\mathrm{\textnormal{#1}}}
\newcommand{\ol}[1]{\overline{#1}}
\newcommand{\ts}[1]{\textsuperscript{#1}}
\newcommand{\lbl}[1]{_{\regtext{#1}}}
\newcommand{\ie}{\textit{i}.\textit{e}.}
\newcommand{\eg}{\textit{e}.\textit{g}.}

\newcommand{\comp}{^{\regtext{C}}}
\newcommand{\card}[1]{\left\vert#1\right\vert}
\newcommand{\interior}[1]{\regtext{int}\!\left(#1\right)}
\newcommand{\proj}{\regtext{proj}}
\newcommand{\norm}[1]{\left\Vert#1\right\Vert}
\newcommand{\floor}[1]{\left\lfloor#1\right\rfloor}
\newcommand{\ceil}[1]{\left\lceil#1\right\rceil}
\newcommand{\abs}[1]{\left\vert#1\right\vert}
\newcommand{\pow}[1]{\regtext{pow}\!\left(#1\right)}
\newcommand{\diag}[1]{\regtext{diag}\!\left(#1\right)}
\newcommand{\eig}[1]{\regtext{eig}\!\left(#1\right)}
\newcommand{\union}{\bigcup}
\newcommand{\intersection}{\bigcap}
\newcommand{\trans}{^\intercal}
\newcommand{\inv}{^{-1}}
\newcommand{\pinv}{^{\dagger}}
\newcommand{\sign}{\regtext{sign}}
\newcommand{\expm}{\regtext{exp}}
\newcommand{\logm}{\regtext{log}}
\newcommand{\skw}{_{\times}}
\newcommand{\bigO}{\mc{O}}
\newcommand{\bdry}[1]{{\partial #1}}
\renewcommand{\ker}[1]{\regtext{ker}\!\left(#1\right)}
\newcommand{\convhull}[1]{\regtext{CH}\!\left(#1\right)}
\newcommand{\dist}[1]{d_2\!\left(#1\right)}
\newcommand{\len}[1]{\left\Vert#1\right\Vert_2}
\newcommand{\trace}{\regtext{trace}}
\newcommand{\degrees}[1]{$#1^\circ$}
\newcommand{\probability}[1]{\Pr\left\lbrace#1\right\rbrace}
\newcommand{\condprob}[2]{\probability{#1 \; | \; #2}}
\newcommand{\set}[1]{\left\lbrace #1 \right\rbrace}
\newcommand{\condset}[2]{\set{#1 \; : \; #2}}

\newcommand{\emptyarr}{[\ ]}
\newcommand{\zeros}{{0}}
\newcommand{\ones}{{1}}
\newcommand{\eye}{{I}}

\newcommand{\normaldist}{\mathcal{N}}
\newcommand{\est}[1]{\hat{#1}}

\newcommand{\state}{\ensuremath{\mathbf{x}}}
\newcommand{\mean}{\mu}
\newcommand{\cov}{\Sigma}
\newcommand{\bearing}{\beta}
\newcommand{\eststate}{\est{\state}}
\newcommand{\mode}{y}
\newcommand{\ctrl}{u}
\newcommand{\noise}{w}
\newcommand{\error}{\varepsilon}
\newcommand{\position}{\mathbf{p}}
\newcommand{\course}{\psi}
\newcommand{\speed}{\varepsilon}
\newcommand{\surgespeed}{U}
\newcommand{\vel}{\mathbf{v}}

\newcommand{\trigfunc}[2]{ \ensuremath{ \regtext{#1}\left( #2 \right)}}
\renewcommand{\sin}[1]{\trigfunc{sin}{#1}}
\renewcommand{\cos}[1]{\trigfunc{cos}{#1}}
\renewcommand{\arctan}[1]{\trigfunc{arctan}{#1}}

\newacronym{IBUS}{IBUS}{Integrated Biomass Utilization System}
\newacronym{DTU}{DTU}{Technical University of Denmark}
\newacronym{SIMAC}{SIMAC}{Svendborg International Maritime Academy}
\newacronym{IMO}{IMO}{International Maritime Organization}
\newacronym{EMSA}{EMSA}{European Maritime Safety Agency}

\newacronym{DDV}{DDV}{Degree of Domain Violation}
\newacronym{TDV}{TDV}{Time to Domain Violation}
\newacronym{BB}{BB}{Branch and Bound}

\newacronym{GPS}{GPS}{Global Positioning System}
\newacronym{AIS}{AIS}{Automatic Identification System}
\newacronym{INS}{INS}{Inertial Navigation System}

\newacronym{ASV}{ASV}{Autonomous Surface Vessel}
\newacronym{COLREGs}{COLREGs}{International Regulations for Preventing Collisions at Sea}
\newacronym{ECDIS}{ECDIS}{Electronic Chart Display and Information System}
\newacronym{ENC}{ENC}{Electronic Navigational Chart}
\newacronym{COLAV}{COLAV}{Collision Avoidance}
\newacronym{CPA}{CPA}{Closest Point of Approach}
\newacronym{TCPA}{TCPA}{Time to Closest Point of Approach}
\newacronym{DCPA}{DCPA}{Distance at Closest Point of Approach}
\newacronym{TTW}{TTW}{Time to Next Waypoint}
\newacronym{SOG}{SOG}{Speed Over Ground}
\newacronym{COG}{COG}{Course Over Ground}
\newacronym{ROT}{ROT}{Rate of Turn}
\newacronym{CR}{CR}{Collision Risk}
\newacronym{OS}{OS}{Ownship}
\newacronym{TV}{TV}{Target Vessel}
\newacronym{OOW}{OOW}{Officer of the Watch}
\newacronym{STCW}{STCW}{International Convention on Standards of Training, Certification and Watchkeeping for Seafarers}
\newacronym{VCS}{VCS}{Voyage Control System}

\newacronym{ATC}{ATC}{Air Traffic Control}
\newacronym{VTS}{VTS}{Vessel Traffic Service}
\newacronym{DP}{DP}{Dynamic Positioning}
\newacronym{UNCLOS}{UNCLOS}{UN Convention of Law of the Sea}


\newacronym{AI}{AI}{Artificial Intelligence}
\newacronym{DL}{DL}{Deep Learning}
\newacronym{MC}{MC}{Monte Carlo}
\newacronym{ANN}{ANN}{Artificial Neural Network}
\newacronym{DNN}{DNN}{Deep Neural Network}
\newacronym{RNN}{RNN}{Recurrent Neural Network}
\newacronym{GNN}{GNN}{Graph Neural Network}
\newacronym{LSTM}{LSTM}{Long-Short Term Memory}
\newacronym{cvae}{CVAE}{Conditional Variational Autoencoder}
\newacronym{RL}{RL}{Reinforcement Learning}
\newacronym{GMM}{GMM}{Guassian Mixture Model}
\newacronym{PCA}{PCA}{Principal Component Analysis}

\newacronym{STL}{STL}{Signal Temporal Logic}

\newacronym{KDE}{KDE}{Kernel Density Estimate}
\newacronym{PDF}{PDF}{Probability Density Function}
\newacronym{ISJ}{ISJ}{Improved Sheather-Jones}
\newacronym{ROS}{ROS}{Rate of Swing}
\newacronym{MASS}{MASS}{Maritime Autonomous Surface Ship}
\newacronym{EKF}{EKF}{Extended Kalman Filter}

\newacronym{I/O}{I/O}{Input-Output}
\newacronym{ACC}{ACC}{Adaptive Cruise Control}
\newacronym{MPC}{MPC}{Model Predictive Control}


\title{Codification of Good Seamanship in Complex and Congested Waterways\thanksref{footnoteinfo}} 


\thanks[footnoteinfo]{This research was sponsored by the Den Danske Maritime Fond in AI-Navigator project (grant number 2022-092).  AIS data and electronic navigational charts have been provided by the Danish Geodata Agency.}

\author[First]{Yaqub A. Prabowo} 
\author[First]{Peter N. Hansen} 
\author[First]{Dimitrios Papageorgiou}
\author[First]{Roberto Galeazzi}

\address[First]{Technical University of Denmark, Dept. of Electrical and Photonics Engineering, Automation and Control Group, 2800 Kgs. Lyngby, Denmark (e-mail: \{yaqpr,pnha,dimpa,roga\}@ dtu.dk).}

\begin{abstract}                
This paper presents a novel method to quantify seafarers' good seamanship during navigation scenarios with multi-vessel encounters -- in open and confined waters --, and to compute COLREG's-compliant trajectories for avoiding collision and grounding. 
The quantification of good seamanship requires knowledge about the state of the vessels (position, heading, and speed) and the surrounding sailing environment. Such information is accessible through the AIS system and the electronic nautical chart. 
The proposed method evaluates mutual collision risk by examining domain violations of each vessel, and comparing them to the seaman's actions. This results in a comprehensive metric of good seamanship. As risk free actions are not always possible in the resolution of a potential collision and grounding, the method adopts a branch-and-bound scheme to identify achievable maneuvers that minimize the risk. Further, the dynamic nature of vessel speed in congested scenarios is considered, recognizing potential changes in both own and target vessels' forward speeds. The proposed method is experimentally evaluated using historical AIS data and sea charts of Danish waters. 
This research contributes to the field by providing a more realistic perspective on seamanship in complex maritime environments.
\end{abstract}


\end{frontmatter}

\section{Introduction}

In today's maritime landscape, navigating vessels, especially in densely populated waterways, requires a nuanced understanding of good seamanship. The increasing maritime traffic elevates the risk of encounters and collisions, necessitating comprehensive safety strategies. Although the COLREGs rule highlights the importance of good seamanship, a significant gap remains without a formal definition or quantification of this crucial aspect. Efforts to bridge this gap include studies focused on risk assessment \citep{stankiewicz_quantifying_2020}, collision-free path generation \citep{zhang_interpretable_2023, stankiewicz_primitive-based_2021}, and alarm system design \citep{wang_risk_2023}.

Notably, a study has focused primarily on good seamanship scoring, overlooking the important scenario of multi-vessel encounters, particularly in open water \citep{stankiewicz_quantifying_2020}. The scoring is based on the overall risk associated with a specific action, calculated through methods such as point-wise approaches like \gls{DCPA} and \gls{TCPA} \citep{jeong_risk_2012}, employing ship domain areas \citep{szlapczynski_ship_2021}, or machine learning \citep{ozturk_evaluating_2019}.

\begin{figure}
    \begin{center}
        \includegraphics[width=9.0cm]{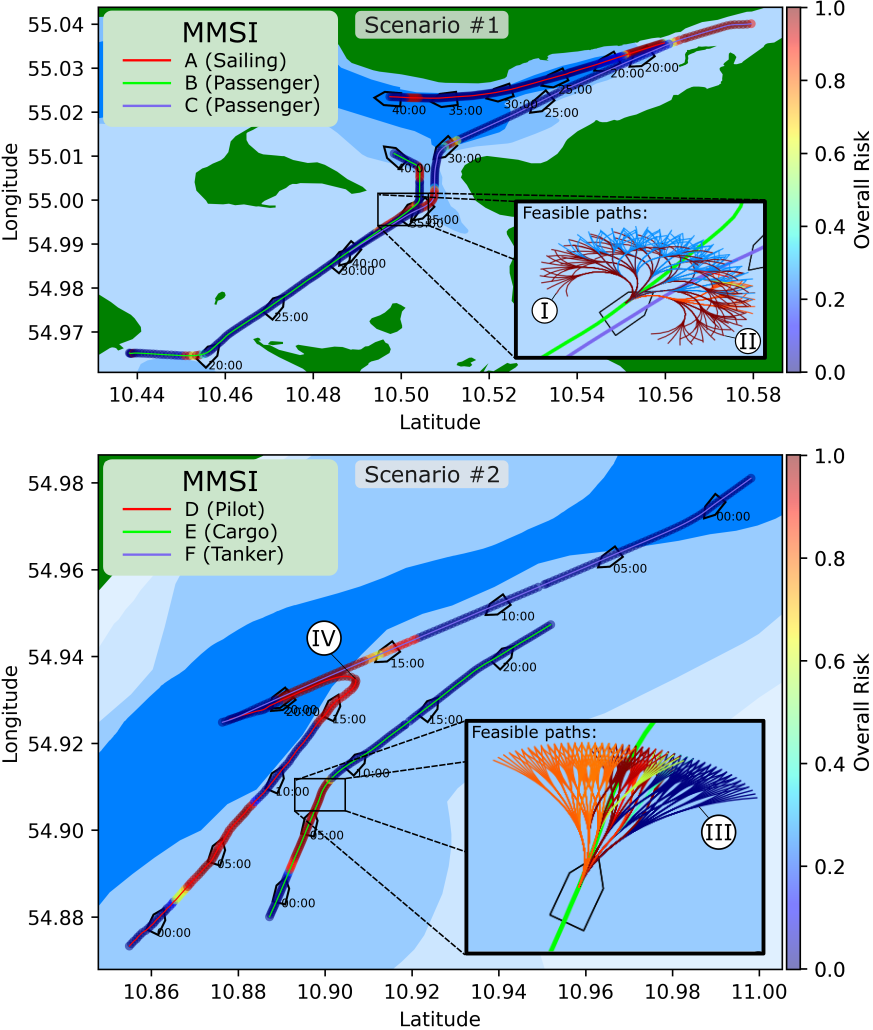}    
        \caption{Scenario \#1: A snapshot from AIS data in the Svendborg area (DK) showcases a complex and congested waterway. The zoomed-in view reveals paths with high grounding risk (I) and high collision risk (II), emphasizing the absence of entirely risk-free routes.
        Scenario \#2: In Langelandsbælt's open waterway (DK), a focused display illustrates risk-free paths (III). A unique situation pointed at (IV) where a Pilot boat following a Tanker causing a high collision risk.} 
        \label{fig:illustration_complex_congested_waterways}
    \end{center}
\end{figure}

A comprehensive approach involves using a ship domain area, rather than the point-wise method, to consider the spatial extent and safety margin of the vessel \citep{szlapczynski_ship_2021}. The ship domain, defined as the designated clear area around the ship, offers a more representative perspective by assessing collision risk through \gls{DDV} and \gls{TDV}.

\begin{figure*}
\begin{center}
\includegraphics[width=17.4cm]{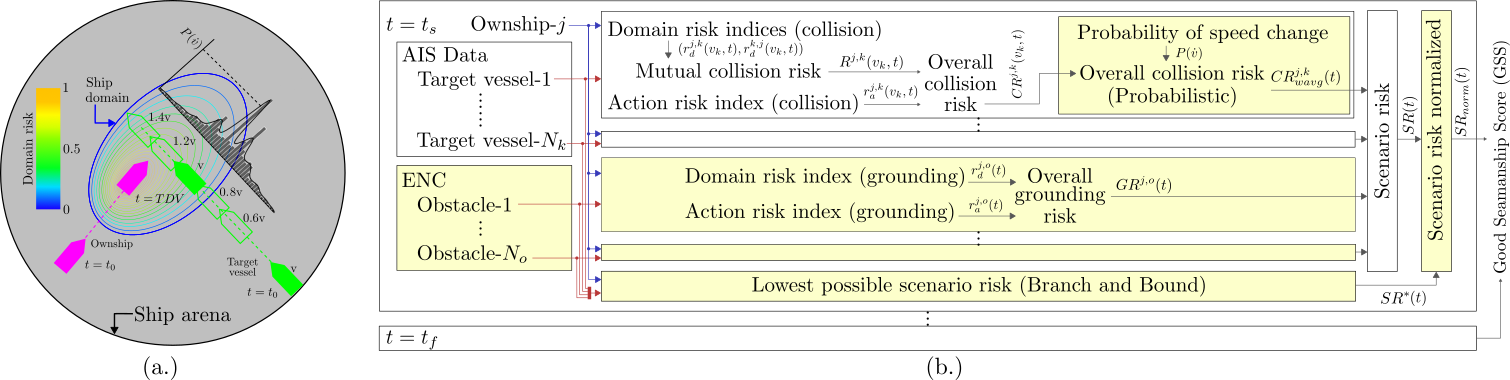}    
\caption{(a.) A scenario of two vessels encounters. (b.) A diagram of how the good seamanship score is quantified.}
\label{fig:diagram_ifac}
\end{center}
\end{figure*}

The \gls{DDV} starts at zero when the target vessel is outside the ownship domain, and it increases as the target vessels enters the ownship domain, following the decentralized ellipse shape \citep{szlapczynski_analysis_2016}. Parameters for this domain can be approximated based on the ship's speed and length \citep{kijima_automatic_2003}.

In narrower areas, as revealed in \citep{hagen_identification_colregs}, the likelihood of collisions and grounding incidents rises. Considering the terrain as static obstacles—landmasses, shallow waters, buoys, and restricted zones—it becomes essential to calculate grounding risk \citep{enevoldsen_grounding-aware_2021}. Grounding risk involves assessing the vessel's bearing angles in shallow waters \citep{zhang_big_nodate}. Another approach, proposed in \citep{bakdi_ais-based_2019} and \citep{bakdi_testbed_2021}, suggests using a polygonal ship domain to calculate the intersection area with the ground shape or the closest horizontal distance to the ground.

Achieving a completely risk-free path may be infeasible, particularly in confined waterways with multiple vessels in transit (shown in Fig. \ref{fig:illustration_complex_congested_waterways}, Scenario \#1 (A) and (B)). It seems unjust to evaluate the seamanship score uniformly when navigators contend with varying levels of complexity in their operating conditions. This study extends the current body of literature by introducing a \gls{BB} method that takes into account vessel kinodynamic constraints. This novel approach provides a more pragmatic means of identifying the lowest risk among the set of potential actions.

A model for control applications, as proposed by \citep{perez_mathematical_2002}, forms the foundation of the vessel dynamics model. However, a limitation arises as most parameters crucial for this model are not readily available through AIS data. Consequently, the kinodynamics constraint is approximated based on the ship's length, which defines the maximum turning radius and subsequently limits maneuverability, as outlined in \citep{hansen_autonomous_2022}.

In \citep{wang2_risk_2023}, AIS data, combined with the ship domain, quantifies collision risk. Recognizing the dynamic nature of the vessel speeds, our study introduces a probabilistic approach to assess collision risk, considering dynamic variations in vessel speeds. Analyzing AIS data with nautical charts reveals speed fluctuations during violations. We employ \gls{PDF} and \gls{KDE} for the initial investigation to represent the probability of speed changes.
To comprehensively validate the scoring of good seamanship, historical AIS data of vessels traversing inner Danish waters and the electronic nautical charts are employed.

The paper is structured as follows: Section \ref{methodology} elaborates the risk assessment framework; Section \ref{result_analysis} presents quantified risks, compares good seamanship scores, analyzes hyperparameters using the \gls{BB} method, and discusses the contribution of the probabilistic approach to overall risk assessment; Sections \ref{discussion} and \ref{conclusion} present the discussion and conclusions.

\section{Methodology} \label{methodology}

\subsection{Preliminary} \label{preliminary}

Utilizing AIS data, a vessel's state at time-$t$, $\boldsymbol{S}(t) = (p_N(t), p_E(t), v(t), \psi(t), L, \mathcal{T}) \in \mathbb{R}^6$, includes the north and east positions ($\boldsymbol{p}(t) = (p_N(t), p_E(t)) \in \mathbb{R}^2$), velocity ($v(t)$), heading ($\psi(t)$), length ($L$) and type of the vessel ($\mathcal{T}$). The good seamanship score ($GSS \in [0,1]$) is quantified using the ownship state $\boldsymbol{S}(t)$ over $t \in [t_s, t_f]$, $N_k$ target vessels and $N_o$ static obstacles. The static obstacle is a set of point $o_i = (o_i^N, o_i^E) \in \mathbb{R}^2$ defined as $O=\{o_1,o_2,..,o_{N_o}\}$. 

The level of violation between ownship-$j$ and a target ship-$k$ is quantified by the collision risk ($CR^{j,k} \in [0,1]$), while the violation between ownship-$j$ and a static obstacle-$o$ is referred to as the grounding risk ($GR^{j,o} \in [0,1]$). These risk values are determined by two key factors: the size of the ship domain ($d$) and the ship arena ($a$) (See Figure~\ref{fig:diagram_ifac}a). Specifically, they correspond to the domain risk index ($r_d \in [0,1]$) and the action risk index ($r_a \in [0,1]$). The ship domain represents the area near the vessel that should remain clear; $r_d$ is relevant when this area is not clear. On the other hand, the ship arena is a designated zone surrounding the vessel where navigators initiate actions to avoid collision or grounding risk; $r_a$ is considered when there is a target vessel or ground within this zone. Consequently, a high collision or grounding risk is indicated if both $r_d$ and $r_a$ are high. The predictability of the navigators is also important to assessed so that the risk in a particular time is quantified over a specific future time horizon $T$.

The future potential for collision can be assessed by assuming a constant velocity for the target vessel. However, a more practical assumption can be made by considering that the target vessel's speed is changing. A probabilistic collision risk ($CR_{\text{wavg}} \in [0,1]$) can be defined as the probability of a change in the target vessel's speed ($P(\dot{v})$). This probabilistic collision risk, along with $GR$, can be integrated into a unified scenario risk ($SR \in [0,1]$).

In high-traffic or narrow channels, it is likely that navigators will face elevated collision or grounding risks, resulting in a very low $GSS$. To ensure fairness, the lowest possible scenario risk ($SR^* \in [0,SR]$) is determined for normalizing $SR$ to $SR_{norm} \in [0,SR]$. $SR_{norm}$ can then be utilized to quantify the $GSS$.
Our contributions are highlighted within the yellow rectangles in Figure~\ref{fig:diagram_ifac}b, representing extensions of the existing method \citep{stankiewicz_quantifying_2020} by incorporating: (1) grounding risk, (2) probabilistic collision risk, and (3) consideration of the lowest possible risk for fairer scoring.

\subsection{Risk Assessment Framework} \label{collision_risk}

\subsubsection{Domain Violation}\label{domain_violation} 

A target vessel may violate the ownship domain at time $t$ equal to \gls{TDV}. The extent of the violation is referred to as \gls{DDV} $\in [0,1]$. As in \citep{szlapczynski_analysis_2016}, the \gls{DDV} value can be derived from a scale factor $f$. Within the context of this paper, the scale factor of vessel-$j$ relative to vessel-$k$ at time-$t$ given the speed of vessel-$k$ and the ship domain $d$ is denoted by $f^{j,k}_d(v_k,t)$. Acquiring $f^{j,k}_d(v_k,t)$ necessitates the imperative inclusion of the relative state between vessel-$j$ and $k$ at time $t$. 

\subsubsection{Domain risk index} 

The domain risk index of an ownship-$j$ given a target vessel $k$ at time $t$ with a constant target vessel-$k$'s speed $v_k$ ($r^{j,k}_d(v_k,t)$) is calculated based on the domain violation between them. It was proposed by \cite{stankiewicz_quantifying_2020} that the domain risk index can be estimated by (\ref{eq:risk_index}) with $\kappa$ and $f_{50}$ as the parameters related to logistic curve.
\begin{equation} \label{eq:risk_index}
r^{j,k}_d(v_k,t) = \frac{1}{1+e^{\kappa(f^{j,k}_d(v_k,t)-f_{50})}}
\end{equation}
\subsubsection{Mutual collision risk}
The risk assessment should extend beyond individual vessels, requiring a comprehensive consideration of mutual risks. Following these principles, the collision risk between vessels $j$ and $k$ at time-$t$ ($R^{j,k}(v_k,t) \in [0,1]$) is calculated as the maximum mutual risk over the specified $T$ (see (\ref{eq:collision_risk})).
\begin{equation} \label{eq:collision_risk}
R^{j,k}(v_k,t) = \max_{\tau \in [t,t+T]} [ r^{j,k}_d(v_k,\tau) \cup r^{k,j}_d(v_k,\tau) ]
\end{equation}
\subsubsection{Overall collision risk} 
The degree of violation may persist (i.e. \gls{DDV} $> 0$) even when the distance between vessels is considerable. The action risk index is calculated using (\ref{eq:risk_index}), with the distinction of employing the ship arena instead of the ship domain. Thus, to obtain the collision risk value ($CR^{j,k}(v_k,t)$), the product of the domain and action risk index is crucial (\ref{eq:overall_collision_risk}), to prevent a persistent high collision risk from distant target vessels.
\begin{equation} \label{eq:overall_collision_risk}
CR^{j,k}(v_k,t) = R^{j,k}(v_k,t) r^{j,k}_a(v_k,t)
\end{equation}
\begin{figure}
    \begin{center}
        \includegraphics[width=8.0cm]{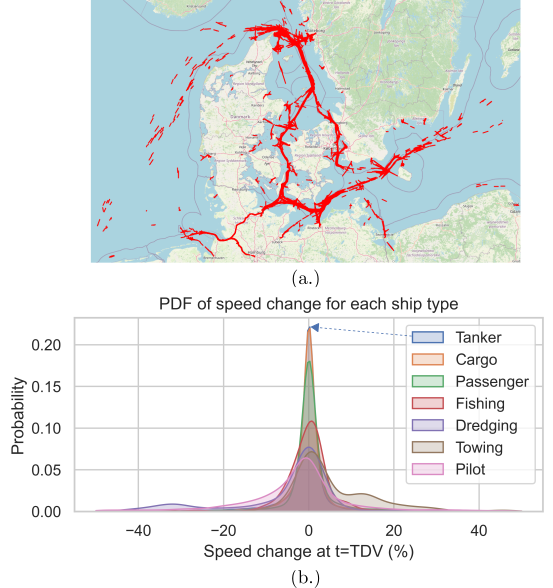}    
        \caption{(a) Scenarios obtained from the AIS data in Denmark on February 1st and 2nd, 2020 (b) The PDF of vessel's speed change at $t=TDV$ grouped by the vessel's type (from 329492 number of violation).} 
        \label{fig:ais_data_and_pdf}
    \end{center}
\end{figure}
\subsubsection{Collision risk under speed change}\label{speed_change_prediction}
Insights can be gained into how navigators adjust vessel speeds in near-collision situations. To accomplish this, the AIS data is filtered to retain scenarios where at least two vessels are likely to collide in the future, determined by \gls{DCPA} below a specified threshold. These critical scenarios are visually depicted in Figure~\ref{fig:ais_data_and_pdf}a.

In assessing risk according to the speed at the time of violation, speed change data is gathered at \gls{TDV} for each identified scenario. The speed change distribution for each vessel is depicted in Figure~\ref{fig:ais_data_and_pdf}b. Tanker and cargo vessels show a near-similar density with the lowest standard deviation, while pilot vessels exhibit the most pronounced variability. Leveraging the \gls{KDE}, it becomes feasible to estimate the probability of speed change during the violation given the vessel type ($P(\dot{v}_{t=\text{TDV}}|\mathcal{T})$).

Recognizing the variability in the target vessel's speed, different collision risk values can arise based on distinct speed changes denoted by $CR^{j,k}(\dot{v},t)$. The final collision risk value is determined by incorporating the probability as a weighting factor, as expressed in Equation (\ref{eq:collision_risk_wavg}). $\dot{v}_{\min}$ and $\dot{v}_{\max}$ represent the minimum and maximum possible speed changes for the vessel.
\begin{equation} \label{eq:collision_risk_wavg}
CR^{j,k}_\text{wavg}(t) =  \frac{\int_{\dot{v}_{\min}}^{\dot{v}_{\max}} P(\dot{v}_{t=TDV}|\mathcal{T}) CR^{j,k}(\dot{v},t) d\dot{v}}{\int_{\dot{v}_{\min}}^{\dot{v}_{\max}} P(\dot{v}_{t=TDV}|\mathcal{T}) d\dot{v}} 
\end{equation}

\subsubsection{Grounding Risk} \label{grounding_risk}

\begin{figure}
    \begin{center}
        \includegraphics[width=6.0cm]{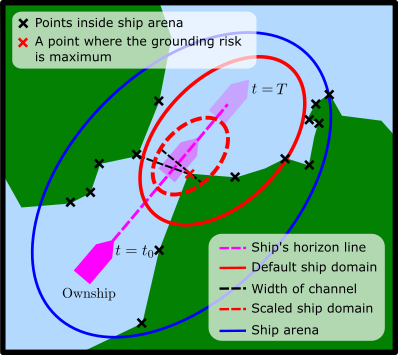}    
        \caption{An illustration demonstrating the quantification of grounding risk. 
        } 
        \label{fig:grounding_risk}
    \end{center}
\end{figure}

Illustrated in Fig. \ref{fig:grounding_risk} is the quantification of grounding risk. Multiple polygons are identified within the seachart, and only $N_o$ points associated with these polygons, located inside the ship arena (depicted as a black cross inside the blue ellipse), are considered. Each obstacle point $o_i$ is assumed as a virtual vessel with zero speed to obtain the domain ($r^{j,o}_d(t)$) and arena ($r^{j,o}_a(t)$) risk index. The grounding risk ($GR^{j,o}(t)$), defined in (\ref{eq:overall_grounding_risk}), is obtained by multiplying these indices.
\begin{equation} \label{eq:overall_grounding_risk}
GR^{j,o}(t) = r^{j,o}_d(t) r^{j,o}_a(t)
\end{equation}
In narrow channels where the width of the ship domain exceeds that of the channel, achieving zero grounding risk is not feasible. To tackle this issue, the ship domain width is proportionally adjusted based on the channel width, as illustrated by the dashed black line. This adjustment is crucial to avoid an excessively high risk of grounding, which could lead to a significantly reduced seamanship score, as demonstrated by the transition from the solid red ellipse to the dashed red ellipse.

\subsubsection{Combining Collision and Grounding Risk} \label{scenario_risk}
The comprehensive assessment of all risks involves recursively uniting them into a scenario risk, as outlined in Algorithm (\ref{alg:scenario_risk}). All target vessels are considered collectively, but only the maximum grounding risk over all static points (indicated by a red cross in Fig. \ref{fig:grounding_risk}) is used. This is done to avoid excessive points that could lead to an unreasonably high overall grounding risk.

\begin{algorithm}
\small
  \caption{Scenario risk for vessel-$j$ towards $N_k$ and $N_o$ number of target vessels and points as the obstacles.}
  \label{alg:scenario_risk}
  \begin{algorithmic}[1]
  \Require $CR^{j,k}_{wavg}(t)$ and $GR^{j,o}(t)$
  \Ensure {Scenario risk ($SR(t)$)}
  \State $SR(t)=0$
  \For{$k := 1 \rightarrow N_k$}
    \State $SR(t)=CR^{j,k}_{wavg}(t)+SR(t)(1-CR^{j,k}_{wavg}(t))$
  \EndFor
  \State $SR(t)=GR^{j,o}_{\max}(t)+SR(t)(1-GR^{j,o}_{\max}(t))$
  \end{algorithmic}
\end{algorithm}

\subsection{Finding the Safest Path from Feasible Actions} \label{most_feasible_action}

\subsubsection{Kinodynamics constraint}

Vessels contend with distinct constraint dynamics, navigating within limits on maneuverability. A pivotal factor influencing a vessel's maneuvering capabilities is its turning radius—a parameter defining the maximum curvature the vessel can achieve while maintaining stability. An approximation proposed by \cite{hansen_autonomous_2022} facilitates the estimation of the maximum turning radius, derived solely from the vessel's length $(L)$. This estimated turning radius plays a crucial role in computing the next position and heading of a vessel at time $t+\Delta t$, given the current velocity, as defined in (\ref{eq:kinodynamics1}) and (\ref{eq:kinodynamics2}).
\begin{equation} \label{eq:kinodynamics1}
\psi(t+\Delta t) = v(t)\Delta t\alpha \delta_{s}(L)
\end{equation}
\begin{equation} \label{eq:kinodynamics2}
\boldsymbol{p}(t+\Delta t) =\boldsymbol{p}(t)
+\frac{\mathcal{R}(\psi(t))}{\alpha \delta_{s}(L)}
\begin{bmatrix}
    \sin{(\psi(t+\Delta t))} \\
    1-\cos{(\psi(t+\Delta t))} \\
\end{bmatrix}
\end{equation}
The rotation matrix ($\mathcal{R}(\psi)$), characterizes the rotational transformation relative to the vessel's heading ($\psi$). The rate of swing is approximated based on the vessel's length $(L)$, denoted as $\delta_{s}(L)$, measured as the angle per unit length. The parameter $\alpha \in [-1,1]$, signifying maximum left and right turns.
\begin{algorithm}
\small
\caption{Branch and Bound}\label{algo:branchandbound}
\begin{algorithmic}[1]
\Require $t, T, N_T,v_{\text{min}}, v_{\text{max}}, N_v, N_\alpha, \boldsymbol{p}(t), \psi(t), SR(t)$
\Ensure $\mathcal{P}^*(t), SR^*(t)$
\State $q(t) \leftarrow (\boldsymbol{p}(t), \psi(t), v(t), SR(t), \text{None})$
\State $Q \leftarrow [q(t)]$
\For {$t_i \in \{ t + \frac{(i-1)\left(T-t\right)}{N_{T}} \mid i = 1, 2, \ldots, N_{T} \}$}
    \State $Q_{\text{temp}} \leftarrow []$
    \While {$Q$ is not empty}
        \State $q(t_i) \leftarrow Q.\text{dequeue()}$
        \For {$v_j \in \{ v_{\min} + \frac{(j-1)\left(v_{\max}-v_{\min}\right)}{N_{v}} \mid j = 1, 2, \ldots, N_{v} \}$}
            \For {$\alpha_k \in \{ -1 + \frac{2(k-1)}{N_{\alpha}} \mid k = 1,2, \ldots, N_{\alpha} \}$}
                \State Get $\psi(t_{i+1})$, $p(t_{i+1})$ with $\alpha_j$ \Comment{Eq. \eqref{eq:kinodynamics1} and \eqref{eq:kinodynamics2}}
                \State Get $SR(t_{i+1})$ \Comment{Alg. 1}
                \State $v(t_{i+1}) \leftarrow v_j$
                \State $q(t_{i+1}) \leftarrow (\boldsymbol{p}(t_{i+1}), \psi(t_{i+1}), v(t_{i+1}), SR(t_{i+1}), q(t_i))$
                \State $Q_{\text{temp}}.\text{enqueue}(q(t_{i+1}))$
            \EndFor
        \EndFor
    \EndWhile
    \State $Q_{\min} \leftarrow \text{argmin}_{SR(t_{i+1})}(Q_{\text{temp}})$
    \For {$q \in Q_{\min}$}
        \State $Q.\text{enqueue}(q)$
    \EndFor
\EndFor
\State $\mathcal{P}^*(t) \leftarrow \text{ConstructPaths}(Q)$
\State $SR^*(t) \leftarrow \text{GetRisk}(\mathcal{P}^*(t))$
\end{algorithmic}
\end{algorithm}
\subsubsection{Branch and Bound}

In the pursuit of deriving the safest feasible path within the framework of vessel kinodynamics and risk assessment, the \gls{BB} method is utilized. This process initiates with the declaration of the current vessel's state as the root of the tree. From this starting point, multiple branches emerge, each representing a potential trajectory characterized by distinct time horizon, speed and heading changes defined by hyperparameters ($N_T, N_\alpha, N_v$) which are the number of sample of time, heading and velocity. 

The algorithm evaluates every potential trajectory to get the safest path. At each branching point, the risk associated with the specific trajectory is assessed. Following this assessment, the \gls{BB} algorithm prioritizes nodes with the lowest calculated risk for further exploration. This iterative and selective approach refines the search space instead of greedily evaluate all trajectories. 

The detailed steps of this method are outlined in Algorithm \ref{algo:branchandbound}. The algorithm operates over discrete time steps for time horizon $T$, exploring possible states represented by position, heading, velocity, and scenario risk. The tree is systematically branched over a velocity range ($v_{\min}$ to $v_{\max}$) and an heading range (by discretizing $\alpha$). The algorithm maintains a queue of states $Q$ which is structured as tuples, containing information about the position, heading, velocity, risk, and a reference to the parent state (See: line-1 and 12). The element of queue is basically the attribute of a vertex of the tree. The set $Q_{\min}$ consists of elements in $Q$ with the lowest scenario risk. The final result consists of optimal trajectories $\mathcal{P}^*(t)$ and the lowest scenario risk $\mathcal{SR}^*(t)$ at time-$t$, obtained by exploring and pruning the solution space in a systematic manner.

\subsection{Good Seamanship Score}

After obtaining $SR^*(t)$, the normalized scenario risk ($SR_{\text{norm}}(t)$) can be quantified. However, since the risk is non-linear, it is necessary to scale it back to the linear scale factor $f$. To determine the linear scale factor $f$, the inverse of the general risk function is required. The general risk function, given the scale factor $f$, is expressed as:
\begin{equation} \label{eq:general_risk_equation}
r(f) = \frac{1}{1+e^{\kappa (f-f_{50})}} \\
\end{equation}
Using this function, the corresponding $f$ value for a given risk $r$ can be derived to: 
\begin{equation} \label{eq:inverse_general_risk_equation}
f(r) = \frac{1}{\kappa} \ln{\left(\frac{1}{r}-1 \right)} + f_{50}
\end{equation}
Subsequently, the $f$ value is normalized to $f_{\text{norm}}$, and then $SR_{\text{norm}}$ is quantified by: 
\begin{equation} \label{eq:normalized_scale_factor}
f_{\text{norm}} = 1+\frac{f(SR)-f(SR^*)}{1-f(SR^*)} \\
\end{equation}
\begin{equation} \label{eq:normalized_risk}
SR_{\text{norm}} = r(f_{\text{norm}})
\end{equation}
The maximum scenario risk $SR_{\max}$ over the specified time period is determined as:
\begin{equation} \label{eq:max_scenario_risk}
SR_{\max} = \max_{t \in [t_s,t_f]} SR_{\text{norm}}(t)
\end{equation}
Finally, $SR_{\max}$ can then be used to get $GSS$ by using Eq. \ref{eq:max_score} to Eq. \ref{eq:good_seamanship_score} that was proposed by \cite{stankiewicz_quantifying_2020}. The $GSS$ is obtained by considering both the maximum score ($J_M$) and cumulative scores ($J_C$) along with a weighting factor ($\beta$). 
\begin{equation} \label{eq:max_score}
J_M =1-SR_{\max}
\end{equation}
\begin{equation} \label{eq:cumulative_score}
J_C =1- \frac{1}{SR_{\max}} \int_{t_s}^{t_f} SR_{norm}(t)dt
\end{equation}
\begin{equation} \label{eq:good_seamanship_score}
GSS = J_M (1+\beta(2 J_C-1)(1-J_M))
\end{equation}
\section{Study Case Analysis}\label{result_analysis}

\begin{figure}
    \begin{center}
        \includegraphics[width=8.5cm]{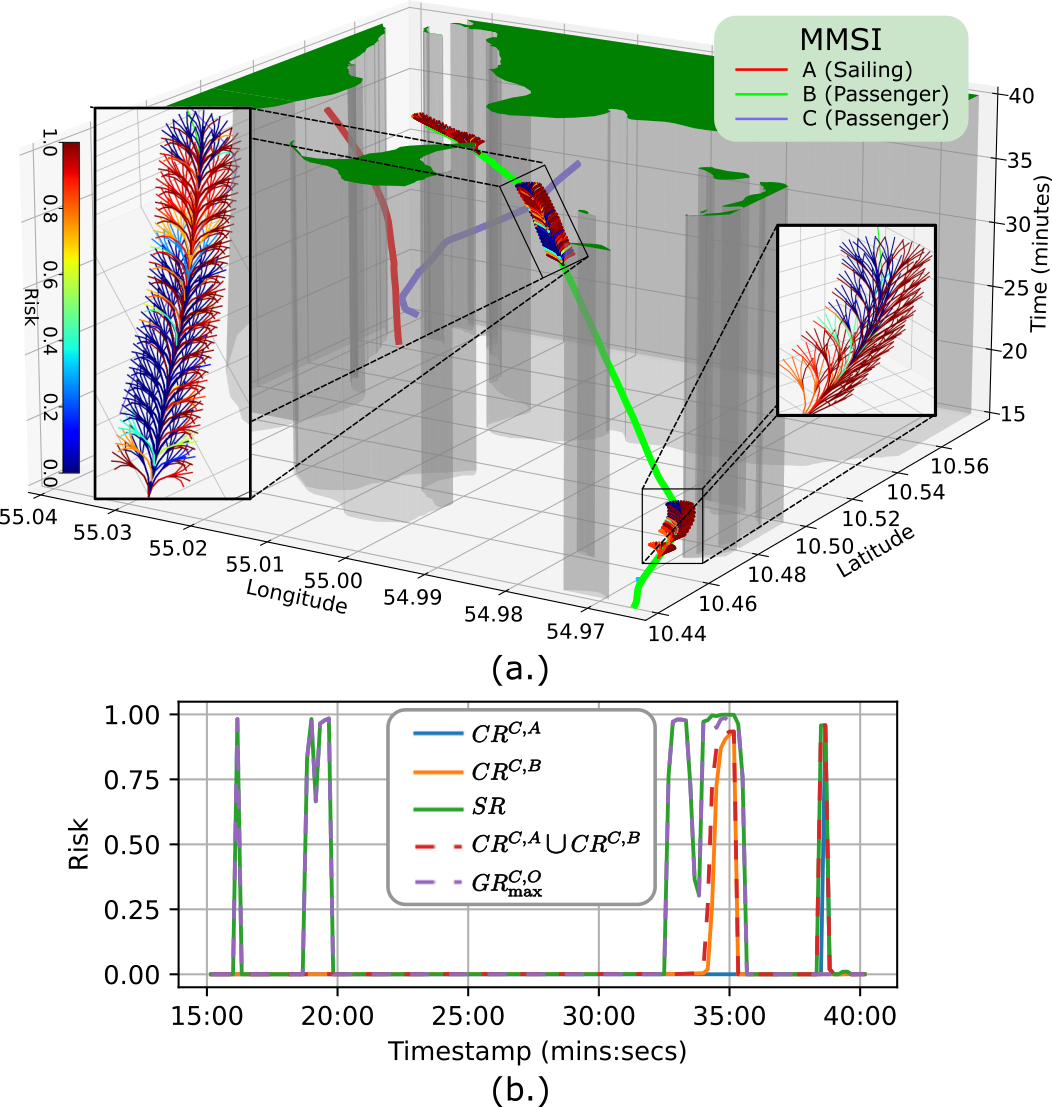}    
        \caption{(a.) Illustration of trees for searching the feasible paths with the lowest risk and (b.) Collision, grounding and scenario risk for vessel C in Scenario \#1.} 
        \label{fig:collision_grounding_scenario_risks}
    \end{center}
\end{figure}

In this paper, two scenarios serve as the study cases, as depicted in Fig. \ref{fig:illustration_complex_congested_waterways}. Scenario \#1 involves vessel encounters in a congested waterways area where the presence of grounding risk is of interest for study. This scenario also leads to situations where achieving zero scenario risk is infeasible. In Scenario \#2, the vessel encounters occur in an open water area. However, a pilot vessel is engaged in pilotage duty, following the tanker vessel, thereby resulting in a high collision risk.

\subsection{Result}

The result was obtained using $T=10$ minutes and diameter of ship arena is 1 NM. Figure \ref{fig:collision_grounding_scenario_risks} illustrates a 3D trajectory plot, where the z-axis corresponds to the timestamp. Concurrently, the associated risk for this trajectory, aligning with Scenario \#1 depicted in Fig. \ref{fig:illustration_complex_congested_waterways}a. Notably, during the period between minute 15 and 20, the vessel encounters a notable increase in grounding risk. Beyond minute 32, the grounding risk remains high, accompanied by a slightly diminished collision risk. Post-minute 38, a brief but pronounced spike in collision risk with another vessel is observed.

\subsection{Branch and Bound}
\begin{figure}
    \begin{center}
        \includegraphics[width=8.5cm]{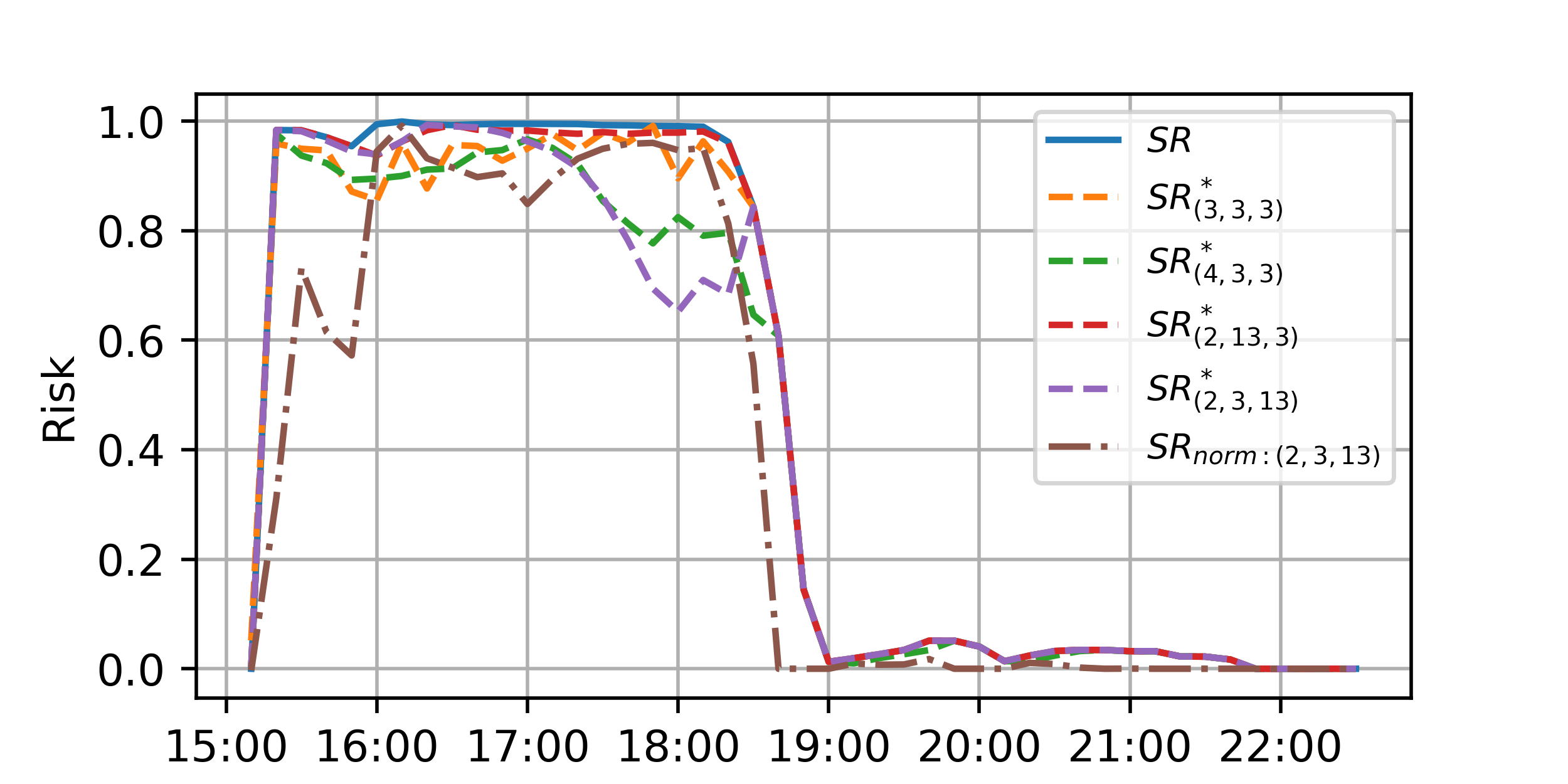}    
        \caption{Lowest possible scenario risk $SR^*_{\left(N_T, N_{\alpha},N_v\right)}$ with different hyperparameters for vessel C in Scenario \#1.} 
        \label{fig:hyperparameters_output}
    \end{center}
\end{figure}

Figure \ref{fig:hyperparameters_output} demonstrates that increasing $N_T$ lowers associated risks for identified paths, while raising $N_{\alpha}$ improves solution exploration. However, higher $N_v$ only slightly enhances path exploration efficiency, with speed range showing less significance compared to heading range. The normalized scenario risk relative to $SR^*_{\left(2,3,13\right)}$ is also shown, representing the risk utilized to quantify the $GSS$.

After extensive hyperparameter tuning, considerable attention was given to choosing a larger value for $N_{\alpha}$ owing to its notable influence on improving path search exploration. Conversely, a lower value for $N_v$ was opted for, as it did not significantly improve performance. The number of time samples, denoted as $N_T$, necessitates careful consideration since an excessively high value would lead to exponential growth in computational costs. This is appropriate with the time complexity for the \gls{BB} algorithm. The best-time complexity is denoted as \(\mathcal{O}(N_{\alpha} N_v N_T)\) but the worst-case is \(\mathcal{O}((N_{\alpha} N_v)^{N_T})\).

\begin{table}[hb]
\begin{center}
\caption{Good seamanship score comparison}\label{tb:gss_comparison}
\begin{tabular}{|c|c|c|c|}
\hline
\multirow{3}{*}{Scenario} & \multirow{3}{*}{Vessel} & \multicolumn{2}{c|}{$GSS$}  \\
\cline{3-4}
 & & \citep{stankiewicz_quantifying_2020} & Proposed \\
 & & +grounding risk &  method \\
\hline
\multirow{3}{*}{\#1} & A & 0.01 & \textbf{0.022} \\
 & B & 0.0058 & \textbf{0.037} \\
 & C & 0.009 & \textbf{0.015} \\ 
\hline
\multirow{3}{*}{\#2} & D & 0.008 & 0.008 \\
 & E & 0.01 & 0.01 \\
 & F & 0.01 & 0.01 \\ 
\hline
\end{tabular}
\end{center}
\end{table}

\subsection{Good Seamanship Score Correction}

Table \ref{tb:gss_comparison} presents a comparison of the good seamanship scores between the methodology from \citep{stankiewicz_quantifying_2020} combined with our grounding risk quantification, and our methodology by normalizing the risk based on the safest possible path. In Scenario \#1, occurring in congested waterways, achieving a completely zero risk is infeasible. Consequently, the score tends to increase as the risk is normalized based on the lowest possible risk. Conversely, in Scenario \#2, set in open waters, although the collision risk is notably high, it is feasible to achieve a state of completely zero risk. As a result, no correction is required in this scenario. Hence, our findings underscore the significant utility of seeking the safest possible path, particularly in congested waterways scenarios.

\subsection{Deterministic vs Probabilistic Risk}

The deterministic collision risk calculation assumes a static speed, leading to an abrupt change in risk. In contrast, the probabilistic approach demonstrates an earlier and more gradual increase in collision risk, as illustrated in Fig. 7, beginning just after minute 0. This smoother escalation is advantageous for collision mitigation in an alarm system, as early risk detection is preferable.

\begin{figure}
    \begin{center}
        \includegraphics[width=8.5cm]{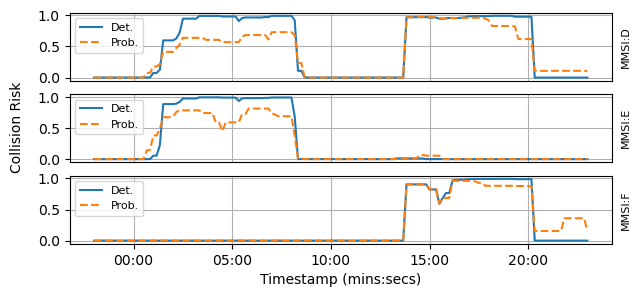}    
        \caption{Comparison between deterministic and probabilistic collision risk in Scenario \#2.} 
        \label{fig:probabilistic_output}
    \end{center}
\end{figure}

Noteworthy, the probabilistic collision risk should consistently be lower or equal to the deterministic risk. Nevertheless, during the transient period after minute 14, both approaches exhibit a similar transient. This alignment is attributed to the sudden left turn of vessel D, a Pilot vessel, which can be observed in Fig. 1b. The vessel's unexpected maneuver results in a sudden spike and continuously high collision risk, as it appears to be following vessel F, a Tanker. This unique scenario may introduce a bias in the risk quantification, prompting a potential research question regarding the appropriate treatment of such situations in terms of risk assessment.

\section{Discussion}\label{discussion}

In our study, a geometric approach is used to score the navigators based on the risk over time. In fact, assessing navigator performance is challenging due to various factors such as the consideration of each vessel mission and priority. For example, in Scenario \#2, our attention turns to a pilot boat swiftly changing course to follow a tanker for a pilotage duty. This prompts the question whether a vessel on special duty should be excluded from the risk assessment or whether a specialized scoring is needed. 
Moreover, future research should thoroughly investigate the dimensions of the ship arena and the time horizon to develop more adaptive solutions, which consider different states and situations of the vessel.

In assessing the risk of collision, attributing accountability to each vessel proves complex. The convergence of two vessels prompts an inquiry into the contributory actions of both or the exclusive responsibility of one. It can be seen in Scenario \#2: when the pilot boat approaches the tanker, the latter experiences a high collision risk as well, however if does not take any action to reduce such risk because of the expected piloting operation. Therefore, considering special vessel mission and their adherence to COLREGs rules are crucial. Combining COLREGs compliance scores with our seamanship scores could potentially lead towards fairer risk assessment.

\section{Conclusions}\label{conclusion}

This research introduced a risk assessment framework designed to quantify a good seamanship score within complex and congested waterways. The integration of collision and grounding risks into an overarching scenario risk, accommodating changes in the target vessel's speed, and normalizing risks against the best feasible path are presented. Utilizing real AIS-data and nautical charts in a case study analysis, the results indicate a more comprehensive risk assessment, enhanced realism through the inclusion of probabilistic collision risk, and improved fairness by acknowledging that not all situations offer risk-free options for navigators.

The application of the \gls{BB} method emerges as a valuable tool for identifying paths with minimized risk and subsequently refining the good seamanship score in an equitable manner. However, acknowledging its basic nature, there is potential for further development into a more efficient tree search method to optimize computation time. Additionally, the incorporation of the probabilistic collision risk opens up opportunities for ongoing investigation, considering factors such as heading changes, and alternative prediction methods beyond KDE such as in \citep{chen_tdv_2023} or \citep{wang_sequential_2022}. These prospects highlight the evolving nature of research in maritime risk assessment and underscore the need for continuous exploration and refinement of methodologies for improved navigational safety.

\begin{ack}
The authors thank Den Danske Maritime Fond for their funding support of the AI Navigator project (grant number 2022-092), enabling the successful completion of this research.
\end{ack}

\bibliography{ifacconf}             

\appendix
\end{document}